\documentclass[twocolumn,showpacs,preprintnumbers,amsmath,amssymb]{revtex4}
\usepackage{tabularx,graphicx}\begin{document}
%\documentstyle[aps]{revtex}
%\documentstyle[preprint,aps]{revtex}
%\begin{document}
\newcommand{\beq}{\begin{equation}}
\newcommand{\eeq}{\end{equation}}
\newcommand{\beqn}{\begin{eqnarray}}
\newcommand{\eeqn}{\end{eqnarray}}
\newcommand{\bmath}{\begin{subequations}}
\newcommand{\emath}{\end{subequations}}

%\newcommand{\bmath}{\begin{mathletters}}
%\newcommand{\emath}{\end{mathletters}}
%\draft
%\twocolumn[\hsize\textwidth\columnwidth\hsize\csname @twocolumnfalse\endcsname 
\title{Spontaneous spinning of a magnet levitating over a superconductor}
\author{J.E. Hirsch$^a$ and D.J. Hirsch}
\address{ $^a$Department of Physics, University of California, San Diego,
La Jolla, CA 92093-0319}

\date{\today} 

\begin{abstract} 
A permanent magnet levitating over a superconductor is found to spontaneously spin, overcoming 
resistance to air friction. We explain the physics behind this remarkable effect.
\end{abstract}
\maketitle 
%\vskip2pc]
It is well known that a  permanent magnet can levitate over a superconductor due to the Meissner effect. Furthermore, for type II 
superconductors, levitation over a flat superconductor is stable due to flux penetration and
pinning of flux lines\cite{hellm}. However it appears not to be widely known that 
under appropriate conditions a levitating magnet has a strong tendency to  start spinning: 
independent of
initial conditions the rotation will reach a terminal velocity and persist, overcoming
resistance due to air friction. After coming across this phenomenon independently we found out
 that this remarkable effect has been seen and discussed before\cite{old1,old2}.
However its physical origin has not been completely elucidated before.
The reader can see examples of this remarkable effect at the website \cite{movie}.

\begin{figure}
%\resizebox{6.5cm}{!}{\includegraphics[width=7cm]{spin1.pdf}}
\resizebox{6.5cm}{!}{\includegraphics[width=7cm]{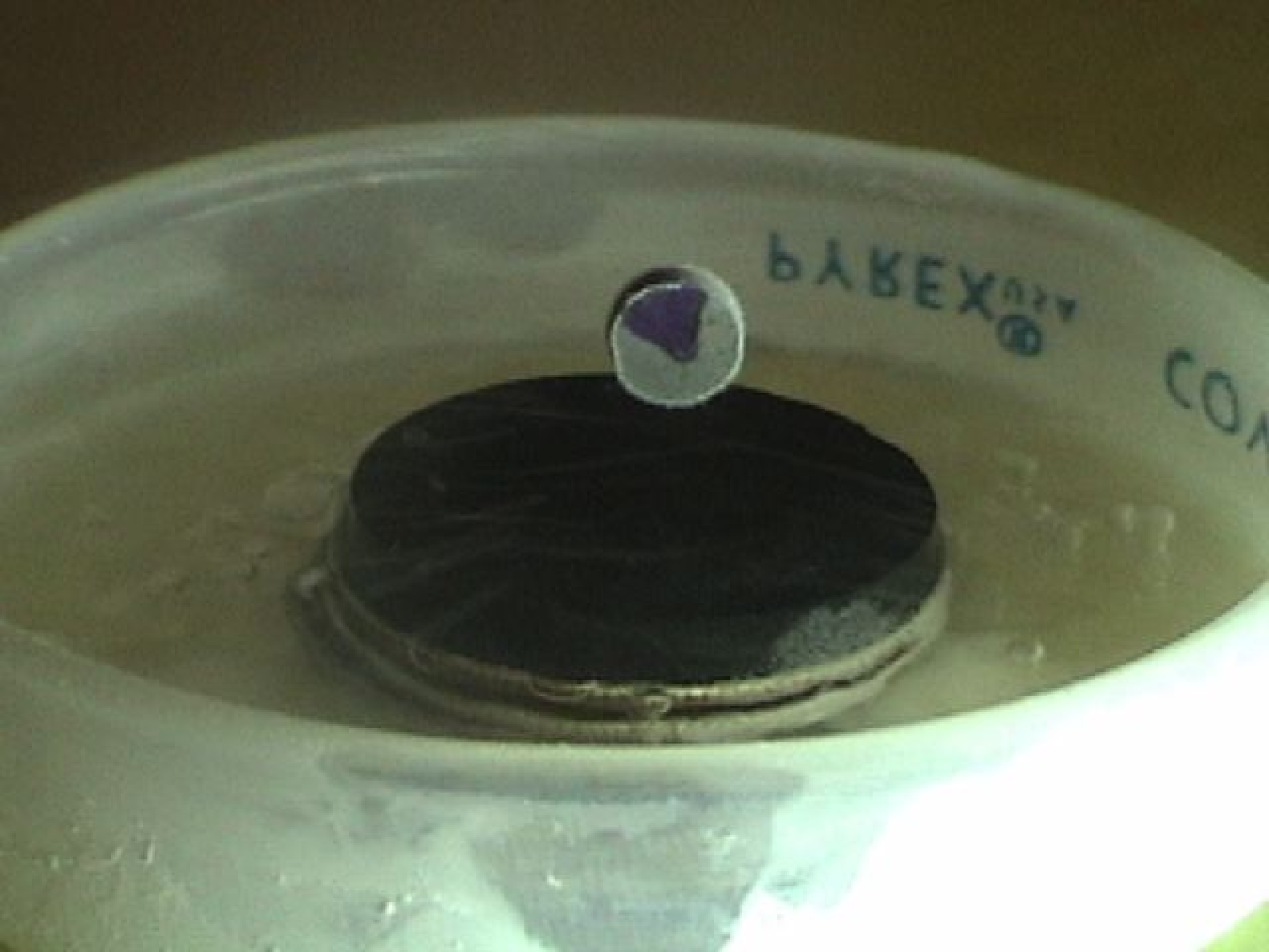}}
\caption{Photograph of experimental setup. A YBCO disk rests on a metal base 
that is submerged in liquid
$N_2$. A $Nd_2Fe_{14}B$ magnet levitates on top of the superconductor.  }
\label{spin1}
\end{figure}

A photograph of our experimental setup is shown in Figure 1. A $YBCO$ disk of diameter $22mm$ and thickness 
$3mm$ rests on a
metal base that is submerged in liquid nitrogen in a flat glass container.
A permanent $Nd_2Fe_{14}B$ magnet disk of radius $R=2.5mm$ and thickness
$h=1.2mm$ is placed vertically over the superconductor, as shown in the figures. 
The magnet is product number $64-1895$ from Radio Shack Corporation.
The height at which the magnet levitates  over the superconductor is approximately $0.4$ mm. It is 
determined by various parameters in the experimental setup such as the thickness of 
the superconductor, the mass of the magnet and the strength of the magnetization,
as discussed in Ref.\cite{hellm}. The magnetization of the permanent magnet is
perpendicular to the magnet disk surface, hence rotation of the magnet around a
horizontal axis going through the disk center does not change the magnetic field.

If the magnet has a high degree of homogeneity it will start rotating almost immediately, while if it is
inhomogeneous it will start rotational oscillations around an equilibrium orientation which increase
in amplitude, until it 'goes over the top' and starts rotating either clockwise or counterclockwise at
random depending on initial conditions. We have found this by  experimentation 
with 8 different magnets, some with a high degree of homogeneity and some with less.
The torque that makes the
magnet  oscillate and spin is to lowest order proportional to the magnet's angular velocity: if the 
magnet is initially at rest with zero angular velocity it is in unstable equilibrium, and any small
perturbation will initiate the motion .

In agreement with previous work\cite{old1,old2} we find that the tendency to spinning is strongest when
the magnet  is situated in a region of large temperature gradient. In the setup of Fig. 1 with container
 height $1.2 cm$ we find that a metal base of approximately $5mm$ is optimal
for strongest spinning tendency. For such height the lower part of the magnet is still inside the
container, while the upper part is outside; this presumably gives rise to largest temperature gradient.
The terminal angular velocity for that case is approximately 4 revolutions/second.
For a lower metal base the tendency to spinning and the terminal velocity is reduced, and
 if we use no metal base in this container, or if we use a deeper container, the 
tendency to spinning disappears. In such cases we would expect the entire magnet to be at a more or less
uniform cold
temperature due to proximity to the liquid $N_2$. Furthermore, we found that with the metal base and
the flat container the spinning can be stopped by placing another container with liquid $N_2$ directly
above the magnet, thus cooling the environment around the upper part of the magnet. Furthermore we
have found that the tendency to spinning is reduced if the magnet is placed not vertically but at an angle;
in that case, the temperature difference between top and bottom of the magnet is obviously reduced.

The observed effect is surprising because there is no obvious source for the kinetic
energy of rotation that the magnet acquires, or for the torque that makes the magnet spin
overcoming air friction.   The effect does not depend on
details of the experimental setup described above. We have observed it
also with other superconducting samples of different shape, even with several
small pieces of superconductor supporting the levitating magnet. The effect is also
seen when several of the magnets described above are put together making a
cylinder of the same radius but larger height.

It was found in reference \cite{old1} that the effect does not originate in
convective currents, i.e. 'wind', originating in the evaporating $N_2$, by placing the magnet inside 
a test tube and observing that the effect persists unchanged. We have repeated the same check. Additionally
it was shown in ref. \cite{old2} that the effect persists when the air pressure is reduced.
Furthermore it was conjectured in refs. \cite{old1,old2} that the effect depends on the thermomagnetic
properties of the magnet, namely on the reversible change of magnetization with temperature.
However no direct experimental evidence was presented that the effect is connected to the magnetism of
the magnet or the superconductivity of the superconductor. In principle one could imagine that the
effect would occur also for a non-magnetic disk in a large temperature gradient if friction with the
environment can be minimized. 

However we have repeated the 
experiment with  superconductors of different thicknesses, and observed a stronger
tendency for spinning and a larger terminal angular velocity under the same conditions of temperature
gradient for increasing thickness. This establishes that the superconductor plays a crucial role beyond simply levitating the
magnet so that it can rotate without significant friction.
Furthermore we have repeated the experiment with three identical magnets forming a cylinder of the same
radius and three times the original height, and found that the tendency to spinning is only slightly reduced;
when we then replaced the middle magnet with another magnet that had been previously heated to a temperature
high enough to completely eliminate its magnetization, it was  found that the tendency to spinning was now 
very significantly reduced. This establishes that the magnetism of the magnet plays a crucial role beyond
simply levitating the magnet and that the spinning effect is larger for larger magnetization.

We have also observed that the effect  depends on the thermal conductivity of the magnet. 
We did not have other magnets of different thermal conductivity but otherwise similar
characteristics available, however, we observed that if the magnet is wrapped with thin
Al foil, the effect becomes weaker or disappears altogether. The thermal conductivity
of $Al$ is much larger than that of $Nd_2Fe_{14}B$. This observation suggests that
the effect will only occur if the thermal conductivity of the magnet is not too large.

Because the effect appears to persists indefinitely it does not seem possible that the 
energy to sustain the motion is supplied by the superconductor. Hence we conclude that it
originates in heat from the environment, i.e. that the rotating magnet constitutes a simple heat
engine as depicted in Figure 2. At any instant when the magnet is
rotating, its upper half will be at a lower temperature than its environment, because half
a period earlier it was in a region of much colder temperature. Similarly, its lower
half will be at temperature higher than its environment because it was in a higher
temperature region half a period ago. Hence the upper half will absorb heat from
the environment and the lower half will release heat to the environment. If the heat absorbed by the
upper half is more than the heat released by the lower half, the difference will be  converted into work. 

\begin{figure}
%\resizebox{7.5cm}{!}{\includegraphics[width=7cm]{spin2.pdf}}
\resizebox{7.5cm}{!}{\includegraphics[width=7cm]{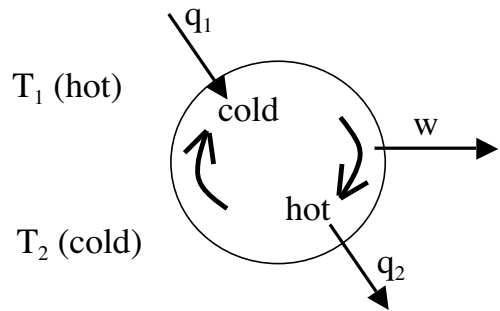}}
\caption{When the magnet rotates, the upper half is colder than its environment and the
lower half is hotter than its environment. Heat $q_1$ flows into the upper half and 
heat $q_2$ flows out of the lower half, the difference is the work $w$ that makes the 
magnet spin. }
\label{spin2}
\end{figure}

It remains to be explained how the conversion of heat to work depicted in Figure 2 occurs.
We argue  that it is  a necessary consequence of the existence of a temperature gradient
in the region where the magnet is 
and of the fact that the magnetization of the $Nd_2Fe_{14}B$ magnet is a decreasing function of temperature
in the temperature range where the magnet is, as discussed in Refs.\cite{old1,old2}.
Qualitatively it works as follows: the superconductor exerts a force on each element of the magnet that is
an increasing function of its magnetization. The magnetization in turn is an increasing function of temperature.
An element of the magnet moving down has a higher temperature than its symmetric counterpart at the
same distance from the superconductor that is moving up. The unequal force exerted by the superconductor on these two magnet elements leads to a net torque
in the same direction as the existing angular velocity.

To make this argument concrete, consider the expression for the energy cost of putting 
vortices originating in a magnetic moment $\vec{m}$ (parallel to the superconductor) through
the superconductor, given in Ref. \cite{hellm}
\beq
E=\frac{H_{c1}L}{\pi}\frac{m}{d}  \equiv C\frac{m}{d}
\eeq
with $H_{c1}$ the lower critical field of the superconductor, $L$ the thickness of the superconductor 
and $d$ the distance from the
superconductor to the magnetic moment. Let $(r,\varphi)$ be the radial and angular coordinates of an
element of magnetic moment $\vec{dm}$ for which $d=d_0+rcos\varphi$, with $d_0$ the distance from
the superconductor to the center of the magnet disk. We measure $\varphi$ with respect to a vertical 
axis $z$. The torque on the magnet due to the force exerted by the superconductor on this element of
magnetic moment is
\beq
d\tau=\frac{ \partial E }{\partial \varphi}= \frac{C\times dm\times  r sin\varphi}{(d_0+r cos\varphi)^2}
\eeq
The element of magnetic moment is given by
\beq
dm=M[T(r,\varphi)] hrdr d\varphi
\eeq
where $M[T]$ is the local magnetization of the disk, which we assume to depend on the local
temperature $T(r,\varphi)$. If the magnet is rotating and has finite thermal conductivity we will have
\beq
T(r,\varphi)\neq T(r,-\varphi)
\eeq
with $T$ lower for the 'rising' side, hence there will be a net torque on the magnet. It is plausible to 
assume to lowest order
\beq
T(r,\varphi)-T(r,-\varphi)=k r sin\varphi
\eeq
with $k$ a constant, hence the net torque is
\beq
\tau=C h k \frac{\partial M}{\partial T} \int_0^Rdr\int_0^\pi d\varphi \frac{r^3 sin^2\varphi}
{(d_0+r cos\varphi)^2}
\eeq
The constant $k$ will depend on the angular velocity $\omega$ of the magnet and its thermal
conductivity $\kappa$, and in particular will go to zero both when $\omega \rightarrow 0$ and 
when $\kappa \rightarrow \infty$. The torque $\vec{\tau}$ is a vector parallel or antiparallel
to the angular velocity $\vec{\omega}$ depending on whether 
$\frac{\partial M}{\partial T}>0$ or $\frac{\partial M}{\partial T}<0$ respectively. In the second case the
torque opposes the motion and together with air friction will rapidly dampen any initial oscillation or
rotation; in the first case instead, the torque will act in the presence of any small initial perturbation
to increase the motion and the magnet will gain angular momentum and kinetic energy. 
In the presence of air friction the magnet will then reach a terminal angular velocity and continue
its rotation indefinitely driven by this torque.

The height $d_0$ at which the magnet levitates is proportional to $L^{1/2}$\cite{hellm}, with $L$ the 
thickness of the superconductor, hence from Eqs. (6) and (1) we expect the torque to increase as
$L^{1/2}$. This is in qualitative agreement with our experimental observations discussed earlier.

The expression Eq. (6) is valid for the case where the entire magnet has magnetization with the same
$\frac{\partial M}{\partial T}$. If as described in our experiment earlier, part of the magnet has no magnetization,
this part does not contribute to the torque Eq. (6) but does contribute to the moment of inertia, thus
reducing the tendency to spinning in agreement with our observations discussed earlier.

\begin{figure}
%\resizebox{8.5cm}{!}{\includegraphics[width=9cm]{spin3.pdf}}
\resizebox{8.5cm}{!}{\includegraphics[width=9cm]{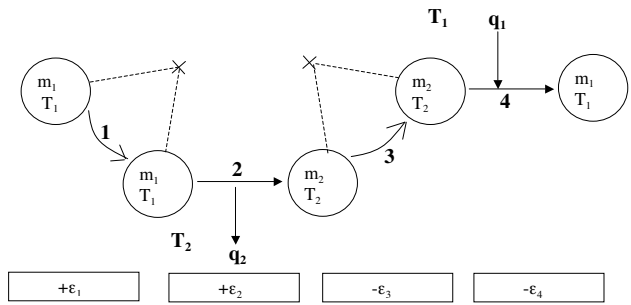}}
\caption{ Four steps of heat engine, labeled {\bf 1} through {\bf 4}. In step {\bf 1} the magnet at temperature
$T_1$ falls from an
initial position at height $d_1$ to a final position at height $d_2$ and the superconductor 
(depicted in the lower part of the figure) gains
energy $\epsilon_1$; in step {\bf 2} , the magnet releases heat $q_2$ and comes to thermal equilibrium with its
environment at a lower temperature $T_2$, and the superconductor gains 
energy $\epsilon_2$; in step {\bf 3}  the magnet rises and the 
superconductor loses energy $\epsilon_3$; in step {\bf 4}  the magnet comes to thermal equilibrium with its
environment at higher temperature $T_1$ by absorbing heat $q_1$ and the superconductor loses 
energy $\epsilon_4$.}
\label{spin3}
\end{figure}

Next we discuss the principle on which this 'heat engine' is based. We consider the simplified example
shown in Figure 3 of a magnetized 'pendulum' on top of a superconductor in the presence of a
temperature gradient. The magnet has two possible positions, one characterized by height $d_1$ and environment
temperature $T_1$ and the other by $d_2$ and $T_2$, with $d_1>d_2$ and $T_1<T_2$. Furthermore, the
magnet's magnetic moment when it is at temperature $T_i$ is $m_i$, with $m_2>m_1$. In the first step the 
magnet (at temperature $T_1$) falls from $d_1$ to $d_2$ and the superconductor increases its energy by
\beq
\epsilon_1=Cm_1(\frac{1}{d_2}-\frac{1}{d_1})
\eeq
 In the second step at position $d_2$, the magnet comes to thermal equilibrium with its environment
at lower temperature $T_2$ by releasing heat $q_2$ , its magnetization increasing to $m_2$. The energy of the superconductor increases by
\beq
\epsilon_2=\frac{C}{d_2}(m_2-m_1)
\eeq
In the third step, the magnet rises again to position $d_1$ and the superconductor lowers its energy by
\beq
\epsilon_3=Cm_2(\frac{1}{d_2}-\frac{1}{d_1})
\eeq
and in the fourth step the magnet absorbs heat $q_1$ and the magnet heats to temperature $T_1$, and its
magnetization decreases to $m_1$, while the superconductor decreases its energy by
\beq 
\epsilon_2=\frac{C}{d_1}(m_2-m_1)
\eeq
Note that $\epsilon_1+\epsilon_2-\epsilon_3-\epsilon_4=0$ as required by conservation of energy
for the superconductor. The difference in the energy increase of the superconductor when the magnet
falls and its decrease when the magnet rises is the gain in the kinetic energy of the magnet:
\beq
\Delta K=\epsilon_3-\epsilon_1=C(m_2-m_1)(\frac{1}{d_2}-\frac{1}{d_1})
\eeq
and conservation of energy in a cycle requires that
\beq
q_1-q_2=\Delta K
\eeq
hence the magnet absorbs more heat than it releases, and the difference is converted into work,
either increasing the kinetic or potential energy of the magnet or dissipated  in air friction or both.

The process described above is an idealization but describes the essence of the mechanism giving
rise to the spontaneous rotation. It is a necessary condition for the above analysis to be valid
that the magnet disk  does not
conduct heat too fast, so that the side coming down has a higher temperature than the side
going up. The thermal conductivity of 
$Nd_2Fe_{14}B$ is $\kappa=9W/m  K$, and its specific heat is $C=0.5\times 10^7 ergs/gr K$.
Using these values and the dimensions of our magnet we estimate the time it takes for heat to conduct across the magnet
as approximately
\beq
\Delta t \sim \frac{C \mu}{\pi \kappa d}\sim 2.5 sec
\eeq
where $\mu=0.174 gr$ is the mass of the magnet. This is much larger than 
the observed  rotation period, hence we conclude that indeed the temperature profile on the magnet is
not equilibrated as required by the mechanism discussed here. Instead, the thermal
conductivity of $Al$ is approximately $25$ times larger than that of $Nd_2Fe_{14}B$, $\kappa=237W/m  K$.
Hence when wrapped with $Al$ foil the magnet would have to turn faster than $10$ revolutions/second to
prevent thermal equilibrium throughout the magnet and allow this mechanism to operate. The resistance from air friction
would however be much larger for such large angular velocity, which is presumably why the effect
does not occur in that case. 

In summary, we have explored the physics behind the unusual observation that a magnet levitating
over a superconductor is found to spontaneously spin under the right conditions. We have 
explained the phenomenon as constituting a simple heat engine, where heat is absorbed from a 
high temperature source, a smaller amount is released to a lower temperature sink, and the difference is
converted into work through the interaction between the magnet and the superconductor.  
Compared to other simple heat engines such as Carnot's or Stirling's the present
one appears to be simplest, as it has only one moving part, the magnet itself, and all the steps in 
the cycle are combined in a single process, the rotation of the magnet. Possible applications  should be explored.

\acknowledgements
The authors are grateful to  B. Maple for kindly providing the superconducting sample and the liquid $N_2$ used in the experiments.

 \end{document}